\documentclass[a4paper,11pt]{article}
\usepackage{pos}
\usepackage{srcltx}
\usepackage{slashed}
\usepackage{bm}
\usepackage{bbm}
\usepackage{ulem}
\usepackage{times}
\usepackage{amsbsy,amsmath,amsfonts}

\usepackage{amssymb}

\newcommand{\trace}[1]{\ensuremath{\langle #1\rangle}}

\title{An accurate relativistic chiral nucleon-nucleon interaction up to the next-to-next-to-leading order}

\author[a,b]{Jun-Xu Lu}
\author[b,c]{Yang Xiao}
\author[b]{Chun-Xuan Wang}
\author*[b,d,e]{Li-Sheng Geng}

\affiliation[a]{School of Space and Environment, Beihang University, Beijing 102206, China}

\affiliation[b]{School of Physics, Beihang University, Beijing 102206, China}
\affiliation[c]{Université  Paris-Saclay, CNRS/IN2P3, IJCLab, 91405 Orsay, France}

\affiliation[d]{Beijing Key Laboratory of Advanced Nuclear Materials and Physics, Beihang University, Beijing 102206, China}
\affiliation[e]{School of Physics and Microelectronics, Zhengzhou University, Zhengzhou, Henan 450001, China }

\emailAdd{lisheng.geng@buaa.edu.cn}

\abstract{We report on the construction of an  accurate relativistic chiral nucleon-nucleon interaction up to the next-to-next-to-leading (NNLO) order.
We compare the so-obtained neutron-proton phaseshifts with the next-to-next-to-next-to-leading order (N$^3$LO) nonrelativistic ones and we show that up to $T_\mathrm{lab.}=200$ MeV, the relativistic chiral nuclear force can describe the PWA93 phaseshifts and inelasticities as well
as its N$^3$LO nonrelativistic counterparts. As a result, the relativistic chiral nuclear force can be readily used  for relativistic ab initio nuclear structure, reaction, as well astrophysical studies.}

\FullConference{%
  The 10th International Workshop on Chiral Dynamics - CD2021\\
  15-19 November 2021\\
  Online
  }

\begin{document}
\maketitle

\section{Introduction}
The nucleon-nucleon ($NN$) interaction plays an essential role in our microscopic understanding of nuclear physics. 
It is the residual part of the strong interaction, for which the underlying theory is quantum chromodynamics (QCD). Although QCD has been well established, due to its two peculiar features, i.e., color confinement and asymptotic freedom, 
one cannot yet understand low-energy nuclear phenomena directly from QCD. As a result, understanding the strong force holding nucleons together has long
been recognized as one of the most difficult questions in Nature~\cite{Bethe:1953,Ishii:2006ec,Epelbaum:2008ga,Machleidt:2011zz}. Traditionally, various phenomenological methods have been employed to derive
the nucleon-nucleon interaction. 
Between 1990 and 1992, Weinberg~\cite{Weinberg:1990rz} proposed that one can use the low-energy effective theory of QCD, i.e., chiral peturbation theory, to derive the $NN$ interaction. Nowadays the so-called chiral nuclear forces  have been constructed up to the fifth order~\cite{Reinert:2017usi,Entem:2017gor}  and sixth order~\cite{RodriguezEntem:2020jgp}, and reached the level of the most refined phenomenological forces, such as Argonne $\textrm{V}_{18}$~\cite{Wiringa:1994wb} and CD-Bonn~\cite{Machleidt:2000ge}. These standard chiral forces and their variants~\cite{Ekstrom:2013kea,Ekstrom:2015rta} are now widely used in ab initio nuclear structure and reaction studies~\cite{Hammer:2019poc,Drischler:2019xuo}. However, almost all these studies are performed in the nonrelativistic framework.

On the other hand, for nuclear structure studies, covariant density functional theories have been extremely successful~\cite{Meng:2005jv,Niksic:2011sg}. The Dirac-Brucker-Hartree-Fock approach has  also been employed to construct a global nucleon-nucleus optical potential~\cite{Xu:2016wqe}. In recent years, there are renewed interests in further refining the Dirac-Bruckner-Hartree-Fock approaches. Using the time-honored Bonn potential developed almost thirty years ago~\cite{Machleidt:1987hj}, self-consistent studies of finite nuclei and nuclear matter have been performed, yielding promising results~\cite{Shen:2019dls}.  Nonetheless, compared to the remarkable progress achieved in the nonrelativistic framework,  much more remain to be done in the relativistic framework. In addition to further developments of relativistic few- and many-body approaches, one crucial missing piece of information is an accurate relativistic chiral nucleon-nucleon interaction. In this talk, we present the first accurate relativistic chiral nuclenon-nucleon interaction recently constructed up to the next-to-next-to-leading (NNLO) order~\cite{Lu:2021gsb}.

This paper is organized as follows. We first briefly review some of the leading order results, then we explain the essential ingredients in constructing the relativistic chiral nucleon-nucleon interaction up to the next-to-next-to-leading order. We comment on the fits of partial waves of low angular momentum ($J\le2$) and the predictions of peripheral waves, followed by a short summary and outlook.

\section{Brief review of leading order results}

In 2016, we proposed to build a relativistic chiral nucleon-nucleon interaction based on the covariant baryon chiral perturbation theory. To achieve this, we developed a covariant power counting scheme~\cite{Ren:2016jna,Xiao:2018jot}  similar to the extended-on-mass-shell (EOMS) scheme of the one-baryon sector~\cite{Gegelia:1999gf,Fuchs:2003qc,Geng:2013xn}. It is interesting to note that our approach is different from the modified Weinberg approach of the Bochum group~\cite{Epelbaum:2012ua,Epelbaum:2015sha,Behrendt:2016nql,Ren:2022glg}, which starts with the manifestly Lorentz-invariant effective  Lagrangian, performs nonrelativistic expansions, employs time-ordered perturbation theory,  and aims at improving the UV behavior  of the Weinberg approach. In contrast, we keep the complete Dirac spinor of the nucleon and aim to provide the most wanted inputs for relativistic ab initio nuclear structure and reaction studies.

At leading order (LO), the covariant scheme has been successfully employed to study   the nucleon-nucleon scattering~\cite{Ren:2016jna,Ren:2017yvw,Bai:2020yml,Wang:2020myr,Bai:2021uim} and hyperon-nucleon scattering~\cite{Li:2016paq,Li:2016mln,Li:2018tbt,Song:2018qqm,Liu:2020uxi,Song:2021yab,Liu:2022nec}. In particular, We showed that in Ref.~\cite{Ren:2016jna}   a reasonable description of the $J=0,1$ $np$ phaseshifts can already be achieved at leading order, in particularly,  the typical low energy features of the $^1S_0$ partial wave~\cite{Ren:2017yvw}. Regarding renormalization group invariance,  the covariant scheme also exhibits some interesting features. For instance,   the extensively studied $^3P_0$ channel becomes RG invariant in the covariant power counting scheme~\cite{Wang:2020myr}.

\section{Next-to-next-to-leading order results}
To construct an accurate relativistic chiral nucleon-nucleon interaction, three more ingredients are needed: 1) contact $NN$ Lagrangians beyond leading order; 2) relevant pion-nucleon couplings determined in the covariant baryon chiral perturbation theory with the EOMS scheme; and 3) two-pion exchange contributions at both leading order and next-to-leading order. As the pion-nucleon scattering has been extensively studied in the EOMS scheme both in SU(2)~\cite{Alarcon:2011zs,Chen:2012nx} and  SU(3)~\cite{Lu:2018zof}, we  focus here on the construction of contact Lagrangians~\cite{Xiao:2018jot} and the calculation of covariant perturbative two-pion exchange contributions~\cite{Xiao:2020ozd}.

\subsection{Covariant chiral nucleon-nucleon contact Lagrangian up to order $\mathcal{O}(q^4)$}

The general expression of a covariant nucleon-nucleon contact Lagrangian reads,
\begin{eqnarray}\label{eq:expression lag}
\frac{1}{\left(2m\right)^{N_d}} \left(\bar{\psi} i \overleftrightarrow {\partial}^{\alpha} i\overleftrightarrow {\partial}^{\beta} ... \Gamma_A \psi\right)
 \partial^{\lambda} \partial^{\mu} ... \left(\bar{\psi} i \overleftrightarrow {\partial}^{\sigma} i\overleftrightarrow {\partial}^{\tau} ... \Gamma_B \psi\right),
\end{eqnarray}
where $\psi=(\psi_p,\psi_n)^T$ denote the relativistic nucleon  field, $\overleftrightarrow {\partial}^{\alpha} = \overrightarrow{\partial}^{\alpha} - \overleftarrow{\partial}^{\alpha}$, 
where $\overrightarrow{\partial}^\alpha /\overleftarrow{\partial}^\alpha$ refers to the derivative on $\psi/ \bar{\psi}$, and $\Gamma \in \{\mathbbm{1}, \gamma_5, \gamma^{\mu}, \gamma_5 \gamma^{\mu}, \sigma^{\mu\nu},g^{\mu\nu}, \epsilon^{\mu\nu\rho\sigma}\} $.  In the above expression, $N_d$ refers to the number of four-derivatives (including $\overleftrightarrow \partial $ and $\partial$ ) in the Lagrangian, $m$ refers to the nucleon mass in the chiral limit, and the factor $1/(2m)^{N_{d}}$ is introduced to unify the dimension of the contact terms. The Lorentz indices $\alpha, \beta ...$ have to be contracted among themselves to fulfill  Lorentz invariance. 

\begin{table}[h]
\caption{Chiral dimensions and properties of fermion bilinears, derivative operators,  Dirac matrices, and Levi-Civita tensor, under parity ($\mathcal{P}$), charge conjugation ($\mathcal{C}$), and hermitian conjugation (h.c.)  transformations.}
\label{Nucleon bilinears}
\centering
\begin{tabular}{ccccccccc}
 \hline
 \hline
 &  $\mathbbm{1}$    &  $\gamma_5$    &    $\gamma_\mu$     &   $\gamma_5\gamma_\mu$   &   $\sigma_{\mu\nu}$  &   $\epsilon_{\mu\nu\rho\sigma}$ &   $\overleftrightarrow \partial_{\mu}$ & $\partial_{\mu}$\\
 \hline
   $\mathcal{O}$   & $0$ & $1$ & $0$ & $0$ & $0$ & $-$ & $0$ & $1$\\
  $\mathcal{P}$   & $+$ & $-$ & $+$ & $-$ & $+$ & $-$ & $+$ & $+$\\
  $\mathcal{C}$   & $+$ & $+$ & $-$ & $+$ & $-$ & $+$ & $-$ & $+$\\
  h.c.            & $+$ & $-$ & $+$ & $+$ & $+$ & $+$ & $-$ & $+$\\

  \hline
  \hline
\end{tabular}
\end{table}

To construct the chiral Lagrangian, one has to specify a proper power counting. In our present case, we need to specify the chiral dimensions of
all the building blocks. In the covariant case, the power counting is more involved, compared to the nonrelativistic case.
The chiral dimensions and properties of fermion bilinears, derivative operators, Dirac matrices, and Levi-Civita tensor under parity, charge conjugation, and hermitian conjugation transformations are listed in Table~\ref{Nucleon bilinears}. The derivative $\partial$ acting on the whole bilinear is of order $\mathcal{O}(q^1)$, while the derivative $\overleftrightarrow {\partial} $ acting inside a bilinear is of $\mathcal{O}(q^0)$ due to the presence of the nucleon mass,
where $q$ denotes a genetic small quantity, such as the nucleon three momentum or the pion mass. The Dirac matrix $\gamma_5$ is of order $\mathcal{O}(q^1)$ because it mixes the large and small components of the Dirac spinor. The Levi-Civita tensor $\epsilon_{\mu \nu \rho \sigma}$ contracting with $n$ derivatives acting inside a bilinear raises the chiral order by $n-1$. If a derivative $\overleftrightarrow \partial$ is contracted with one of the Dirac matrices $\gamma_5\gamma^{\mu}$ or $\sigma^{\mu\nu}$  in a different bilinear, the matrix element is of $\mathcal{O}(q^1)$,
as can be explicitly checked by means of the equation of motion (EoM). Therefore, at each order in the powering counting, only a finite number of $\partial$ and $\epsilon_{\mu\nu\rho\sigma}$ appear. However, in principle, any number of pairwise contracted $i\overleftrightarrow{\partial}$ of the form
\begin{equation}\label{eq:npartial}
\widetilde{\mathcal{O}}_{\Gamma_A \Gamma_B}^{(n)}=\frac{1}{(2m)^{2n}}\left(\bar{\psi} i \overleftrightarrow{\partial}^{\mu_1} i \overleftrightarrow{\partial}^{\mu_2} ... i \overleftrightarrow{\partial}^{\mu_n} \Gamma_{A}^{\alpha}\psi\right)
 \times\left(\bar{\psi} i \overleftrightarrow{\partial}_{\mu_1} i \overleftrightarrow{\partial}_{\mu_2} ... i \overleftrightarrow{\partial}_{\mu_n} \Gamma_{B \alpha}\psi\right),
\end{equation}
 is allowed, since it is of  $\mathcal{O}(q^0)$.  On the other hand, the structure $
\frac{\left[\left(p_1+p_3\right) \cdot \left(p_2+p_4\right)\right]^{n}}{\left(2m\right)^{2n}}$
can be rewritten as $
\left[1+\frac{\left(s-4m^2\right)-u}{4m^2}\right]^{n},$
with $s-4m^2=-(p_1-p_2)^2=-(p_3-p_4)^2 \sim \mathcal{O}(q^2)$ and $u=(p_1-p_4)^2\sim \mathcal{O}(q^2)$. 
Therefore, at $\mathcal{O}(q^0)$, only the terms with $n=0,1,2$ are needed, at  $\mathcal{O}(q^2)$ only the terms with $n=0,1$ are needed, and at $\mathcal{O}(q^4)$ only the terms with $n=0$ are needed since
no new structures appear for $n$ larger than those specified above.

Following the general principles of constructing effective Lagrangians and guided by Table~\ref{Nucleon bilinears}, one can 
write down all the terms of  $\mathcal{O}(q^0)$, $\mathcal{O}(q^2)$, and $\mathcal{O}(q^4)$. However, not all of them are independent up to the order of our concern
and one can use the EoM  to eliminate the redundant terms.
The EoM for the nucleon refers to the well-known Dirac equation at LO
\begin{equation}
 \slashed{\partial}\psi = \gamma^\mu \partial_\mu \psi = - i m \psi + \mathcal{O}(q) \,,
\end{equation}
and its Hermitian conjugate. Up to higher order corrections one can replace \(\slashed{\partial}\psi\) by \(- i m \psi\) and \(\bar \psi \overleftarrow{\slashed{\partial}}\) by \( i m \bar \psi\).
To fully utilize this EoM,  one needs to transform terms that do not contain $\slashed{\partial}$ into forms containing it. The master  formula is
\begin{equation} \begin{aligned} \label{eq:eomrel}
 - 2i m \left(\bar{\psi} {\Gamma} \psi\right)&\approx  \left(\bar{\psi} {\Gamma^\prime}^\lambda \overleftrightarrow{\partial}_{\lambda}\psi\right)+\partial_{\lambda}\left(\bar{\psi} {\Gamma^{\prime\prime}}^\lambda  \psi\right) \,,
\end{aligned} \end{equation}
where $\Gamma$, $\Gamma^{\prime}$, and $\Gamma^{\prime\prime}$ are Dirac matrices given in Ref.~\cite{Xiao:2018jot} and $\approx$ indicates equal up to higher orders.
Using the EoM together with the decomposition of Dirac matrices, one can obtain a set of linear relations. With these relations,  we obtained a minimal and complete set of relativistic $NN$ contact Lagrangian terms of 40 up to $\mathcal{O}(q^4)$~\cite{Xiao:2018jot}.

\subsection{Two-pion exchange contributions}

In order to calculate the contributions of two-pion exchanges, we need the following  LO and NLO $\pi N$ Lagrangians~\cite{Chen:2012nx},
\begin{align}\label{eq:lagpiN}
\mathcal{L}_{\pi N}^{(1)} &=\bar{\psi}\left( {\rm{i}} \slashed{D}-m + \frac{g_A}{2} \slashed{u} \gamma_5 \right) \psi ,\\
\mathcal{L}_{\pi N}^{(2)} &=c_{1}\trace{\chi_{+}}\bar{\psi}\psi- \frac{c_{2}}{4m^2}\trace{u^{\mu}u^\nu}\left(\bar{\psi}D_{\mu}D_{\nu}\psi +h.c.\right)+\frac{c_3}{2}\trace{u^2}\bar{\psi} \psi-\frac{c_4}{4}\bar{\psi}\gamma^{\mu}\gamma^{\nu}\left[u_{\mu},u_{\nu}\right] \psi,
\end{align}
where the covariant derivative $D_\mu$ is defined as $D_{\mu}=\partial_{\mu}+\Gamma_{\mu}$ with $
\nonumber \Gamma_{\mu}=\frac{1}{2}\left(u^\dag \partial_{\mu} u+ u \partial_{\mu} u^\dag\right)$ and $u={\rm{exp}}\left(\frac{{\rm{i}}\Phi}{2f_\pi}\right)$.
The pion field $\Phi$ is a $2 \times 2$ matrix $
\nonumber\Phi=\left(
 \begin{matrix}
   \pi^0 &  \sqrt{2} \pi^{+}\\
   \sqrt{2} \pi^{-} &  -\pi^0\\
  \end{matrix}
 \right),$
and the axial current type quantity $
 u_{\mu}={\rm{i}}\left(u^\dag \partial_{\mu} u- u \partial_{\mu} u^\dag\right),$
where $\chi_{+}=u^\dag \chi u + u \chi u^\dag$ with $\chi=\mathcal{M}=diag\left(m_\pi^2,m_\pi^2\right)$. The following values for the
  relevant LECs and masses are adopted in the numerical calculation: the pion decay constant $f_{\pi}=92.4$ MeV, the axial coupling constant $g_A=1.29$~\cite{Machleidt:2011zz}, the nucleon mass $m_n=939$ MeV, the pion mass $m_\pi=139$ MeV~\cite{Tanabashi:2018oca}, and the low-energy constants $ c_1 =-1.39$, $c_2=4.01$, $c_3=-6.61$, $c_4=3.92$, all in units of GeV$^{-1}$, taken from Ref.~\cite{Chen:2012nx} .
   
In Ref.~\cite{Xiao:2020ozd}, it was shown that the relativistic effects in the perturbative two-pion-exchange (TPE) contributions do improve the description of the peripheral $NN$ scattering data compared to their nonrelativistic counterparts. In Ref.~\cite{Wang:2021kos}, the same feature is found also for the non-perturbative TPE  contributions.

\subsection{Relativistic chiral nucleon-nucleon interaction}
Due to the non-perturbative nature of the nucleon-nucleon interaction, we need to solve a relativistic scattering equation with the chiral potential as inputs. In this work, we solve the following relativistic Blankenbecler-Sugar (BbS) equation~\cite{Blankenbecler:1965gx}, 
\begin{equation}\label{BbSE}
  T(\bm{p}',\bm{p},s)=V(\bm{p}',\bm{p},s)
  +\int\frac{\mathrm{d}^3\bm{k}}{(2\pi)^3}V(\bm{p}',\bm{k},s)\frac{m^2}{E_k}\frac{1}{\bm{q}_{cm}^2-\bm{k}^2-i\epsilon}T(\bm{k},\bm{p},s),
\end{equation}
where $|\bm{q}_{cm}|=\sqrt{s/4-m^2}$ is the nucleon momentum on the mass shell in the center of mass (c.m.) frame, and a regulator  $f_R(p)=\theta(\Lambda^2-p^2)$ is introduced to regularize the potential. Up to NNLO, the relativistic chiral potential consists of the following terms
\begin{equation}\label{NNForce}
  V=V_{\mathrm{CT}}^{\mathrm{LO}}+V_{\mathrm{CT}}^{\mathrm{NLO}}+V_{\mathrm{OPE}}+V_{\mathrm{TPE}}^{\mathrm{NLO}}+V_{\mathrm{TPE}}^{\mathrm{NNLO}}-V_{\mathrm{IOPE}},
\end{equation}
in which the first two terms refer to the LO [$\mathcal{O}(q^0)$] and NLO [$\mathcal{O}(q^2)$] contact contributions, while the next three terms denote the one-pion exchange (OPE), leading, and subleading TPE contributions. The last term represents the iterated OPE contribution. 

Following the strategy adopted in nonrelativistic studies, e.g., Refs.~\cite{Epelbaum:2014sza,Ordonez:1993tn}, we perform a global fit to the $np$ phaseshifts for all the partial waves with total angular momentum $J\leq 2$~\cite{Stoks:1993tb}. For each partial wave, we choose eight data points with laboratory kinetic energy $T_{\rm{lab}}=1,5,10,25,50,100,150,200$ MeV for the fitting. The $\chi^2$-like function to be minimized, $\tilde{\chi}^2$, is defined as 
\begin{equation}
    \tilde{\chi}^2=\sum_i(\delta^i-\delta^i_{\rm{PWA93}})^2,
\end{equation}
where $\delta^i$ are theoretical phaseshifts or mixing angles, and $\delta^i_\mathrm{PWA93}$ are their empirical PWA93 counterparts~\cite{Stoks:1993tb}. 

\begin{figure*}[htbp]
\centering
\includegraphics[width=0.36\textwidth]{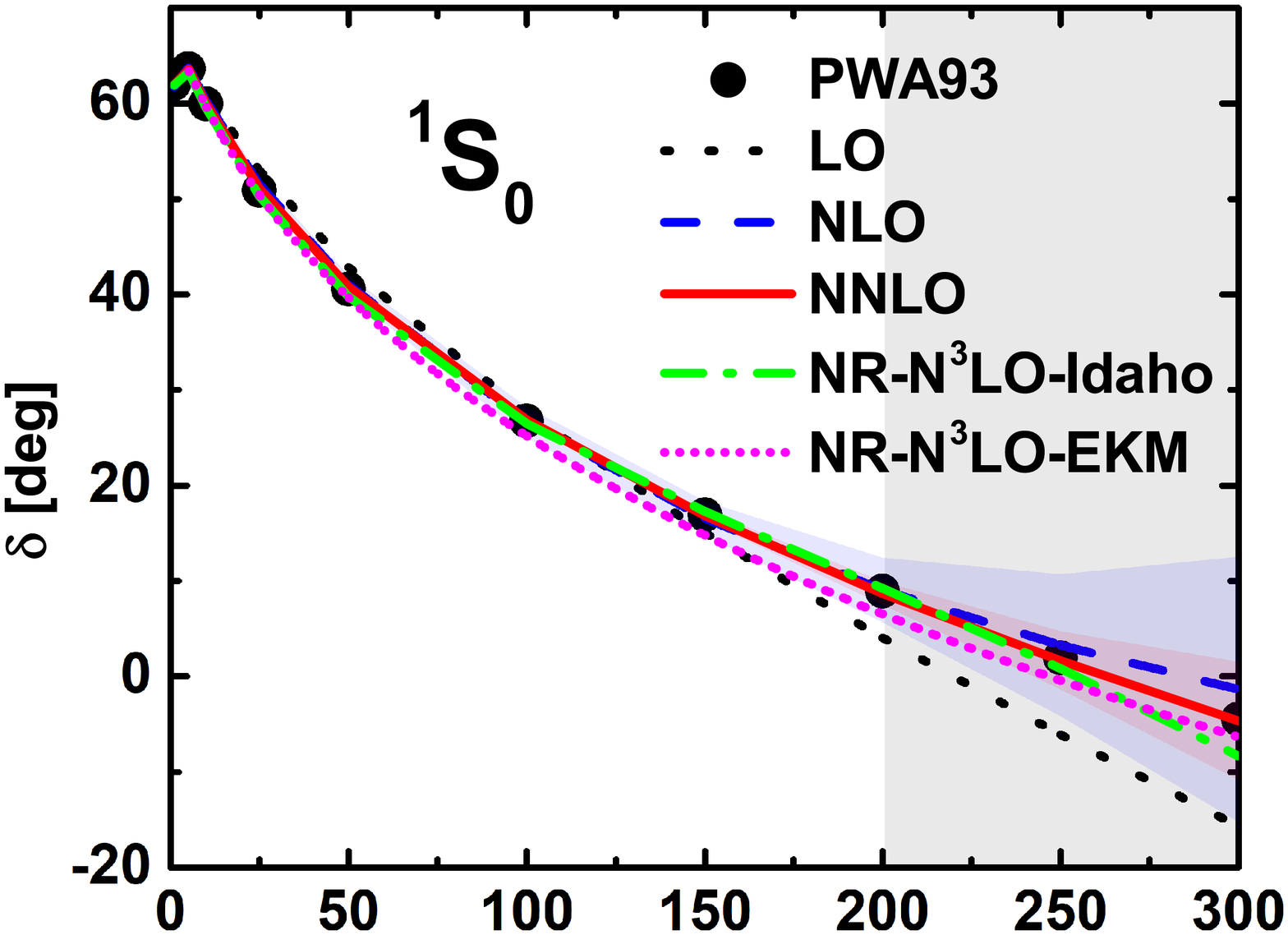}\hspace{-13mm}
\includegraphics[width=0.36\textwidth]{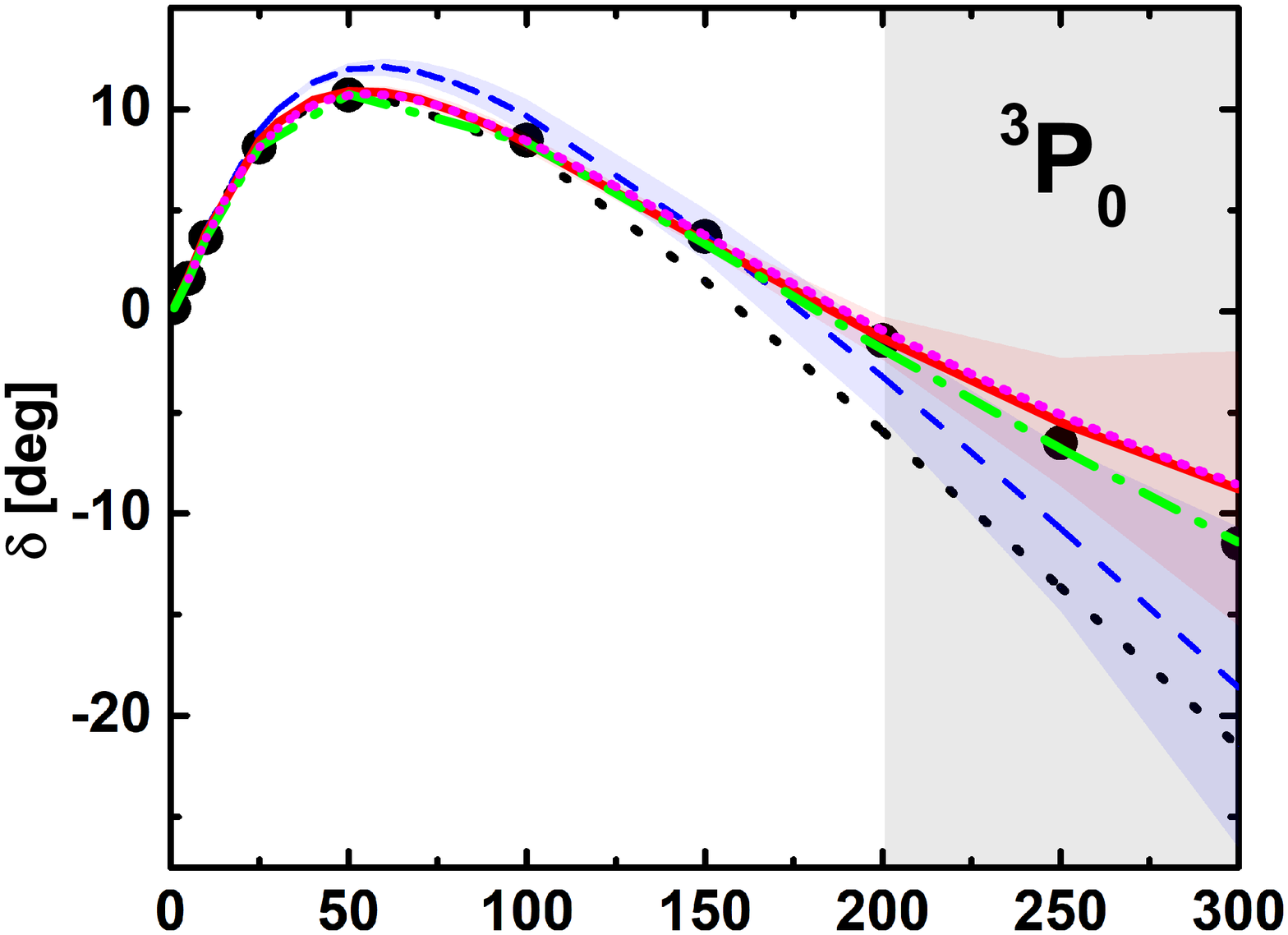}\hspace{-13mm}
\includegraphics[width=0.36\textwidth]{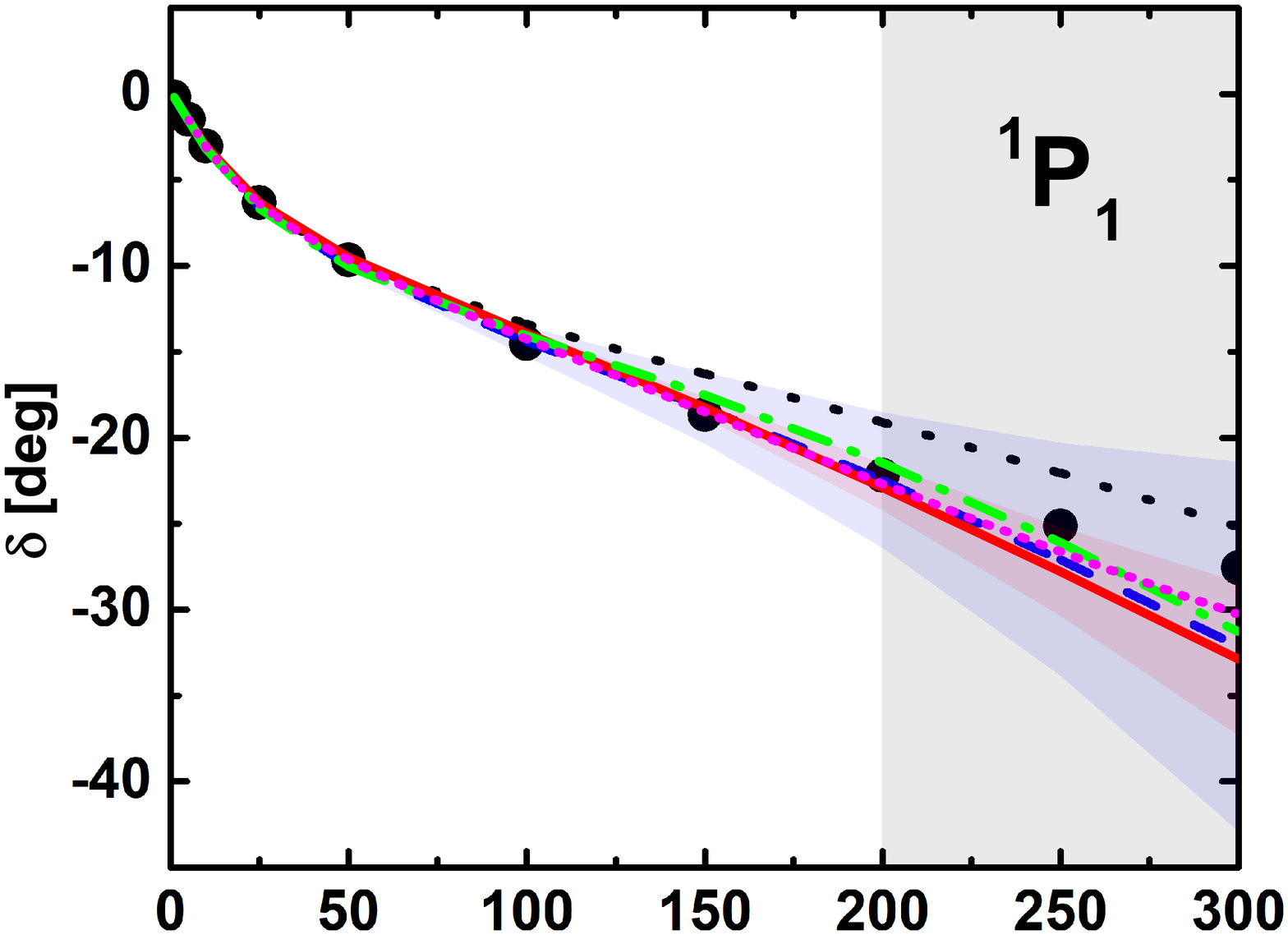}\\ \vspace{-9mm}
\includegraphics[width=0.36\textwidth]{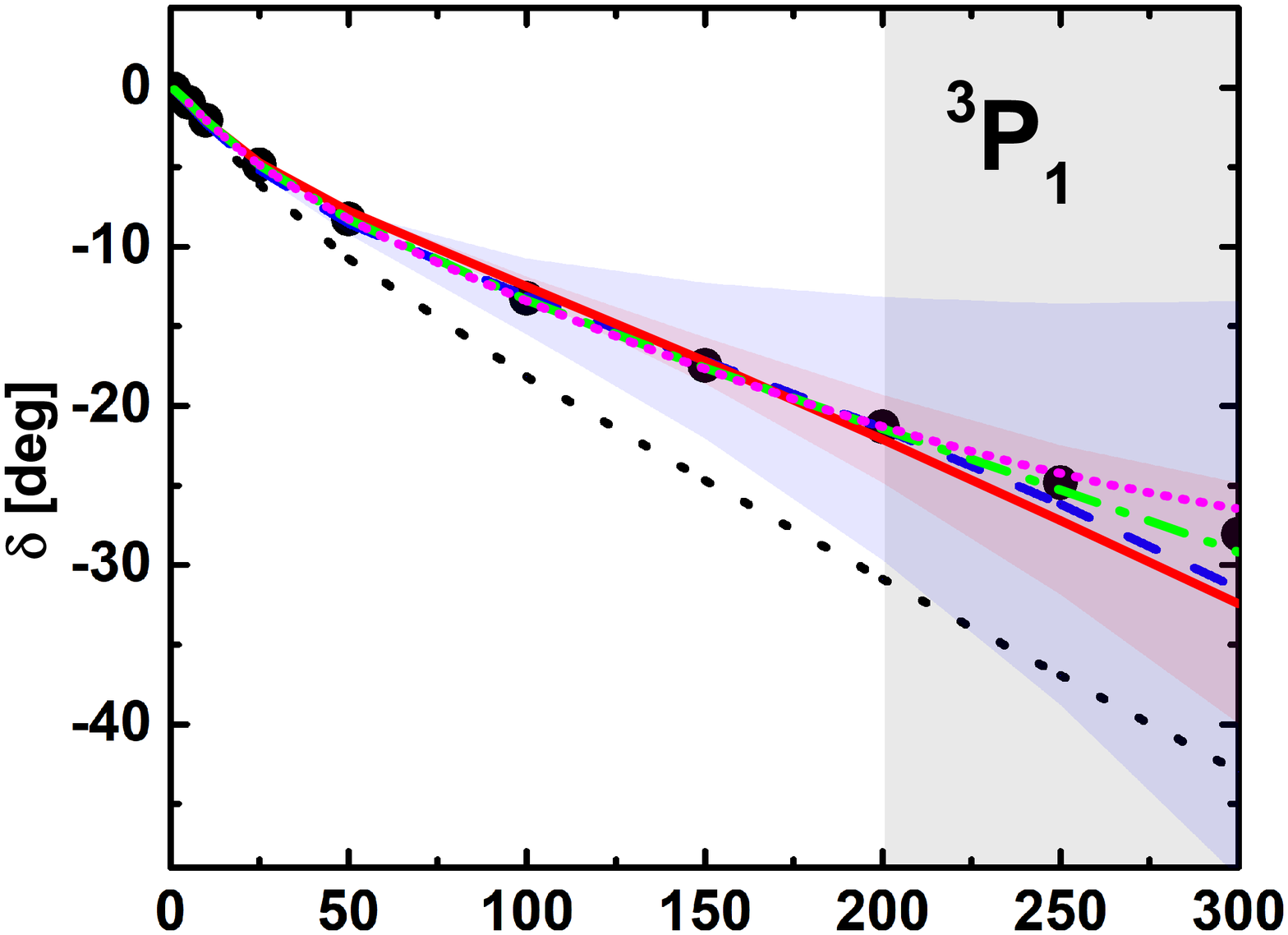}\hspace{-13mm}
\includegraphics[width=0.36\textwidth]{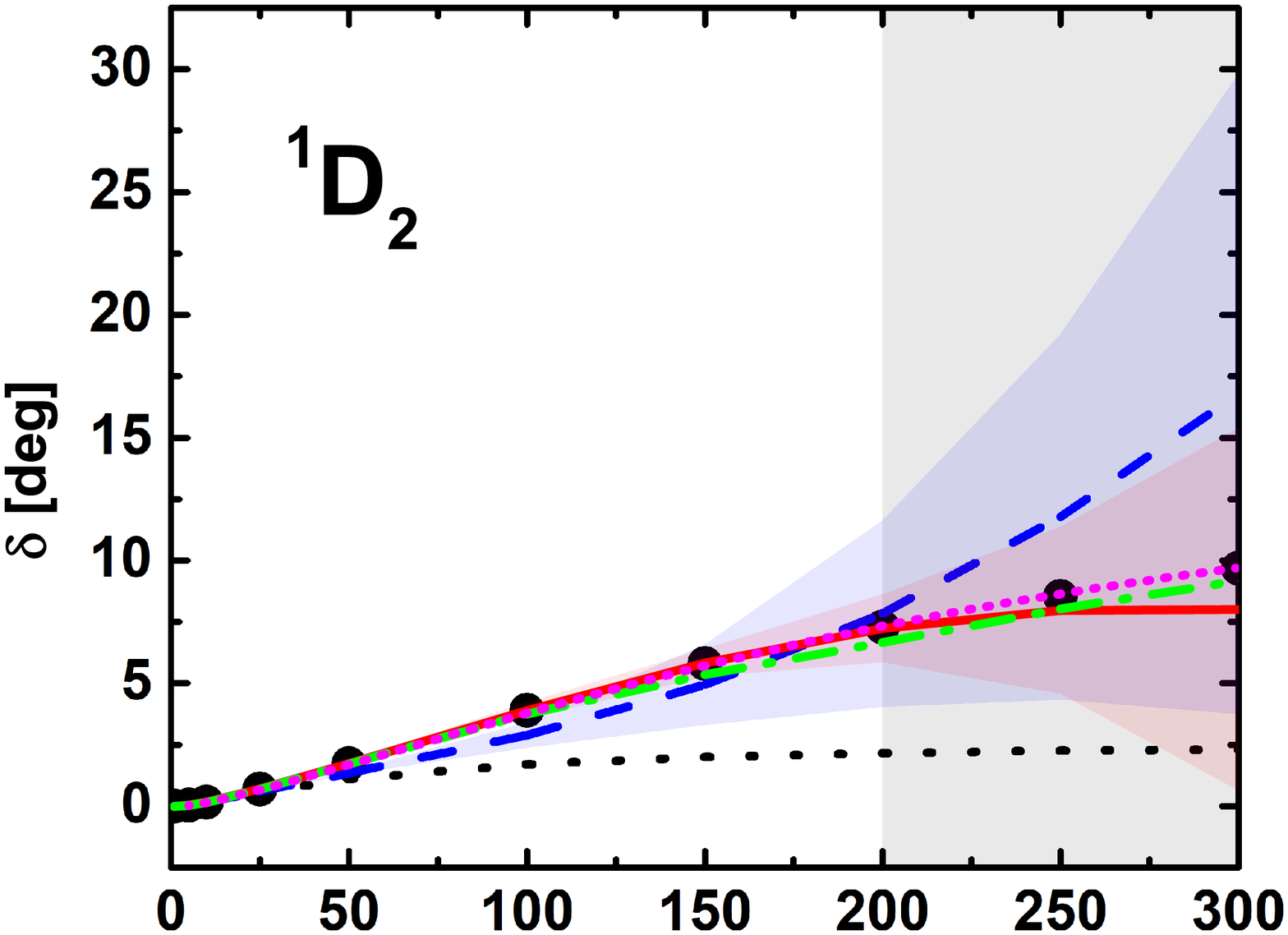}\hspace{-13mm}
\includegraphics[width=0.36\textwidth]{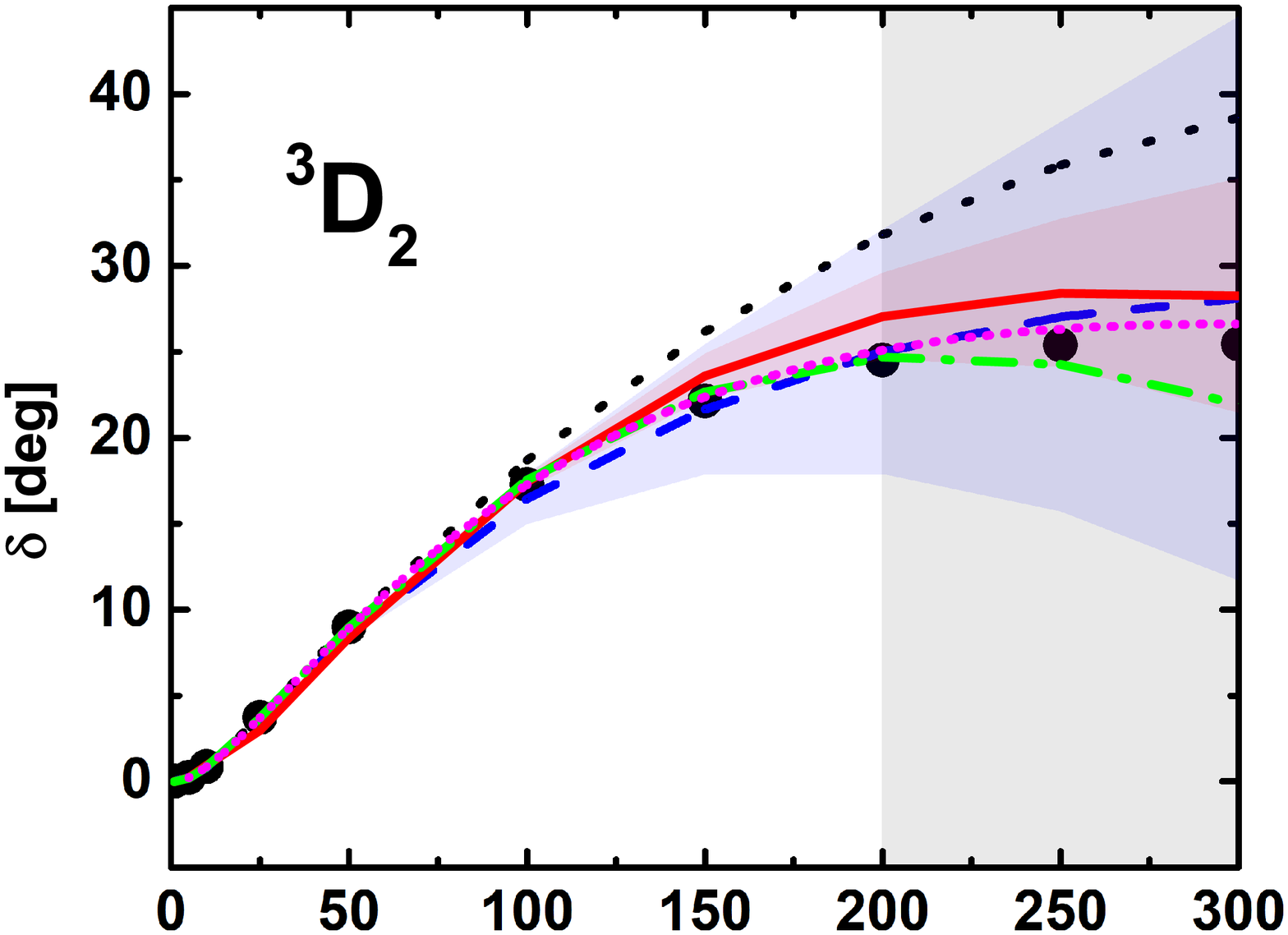}\\ \vspace{-9mm}
\includegraphics[width=0.36\textwidth]{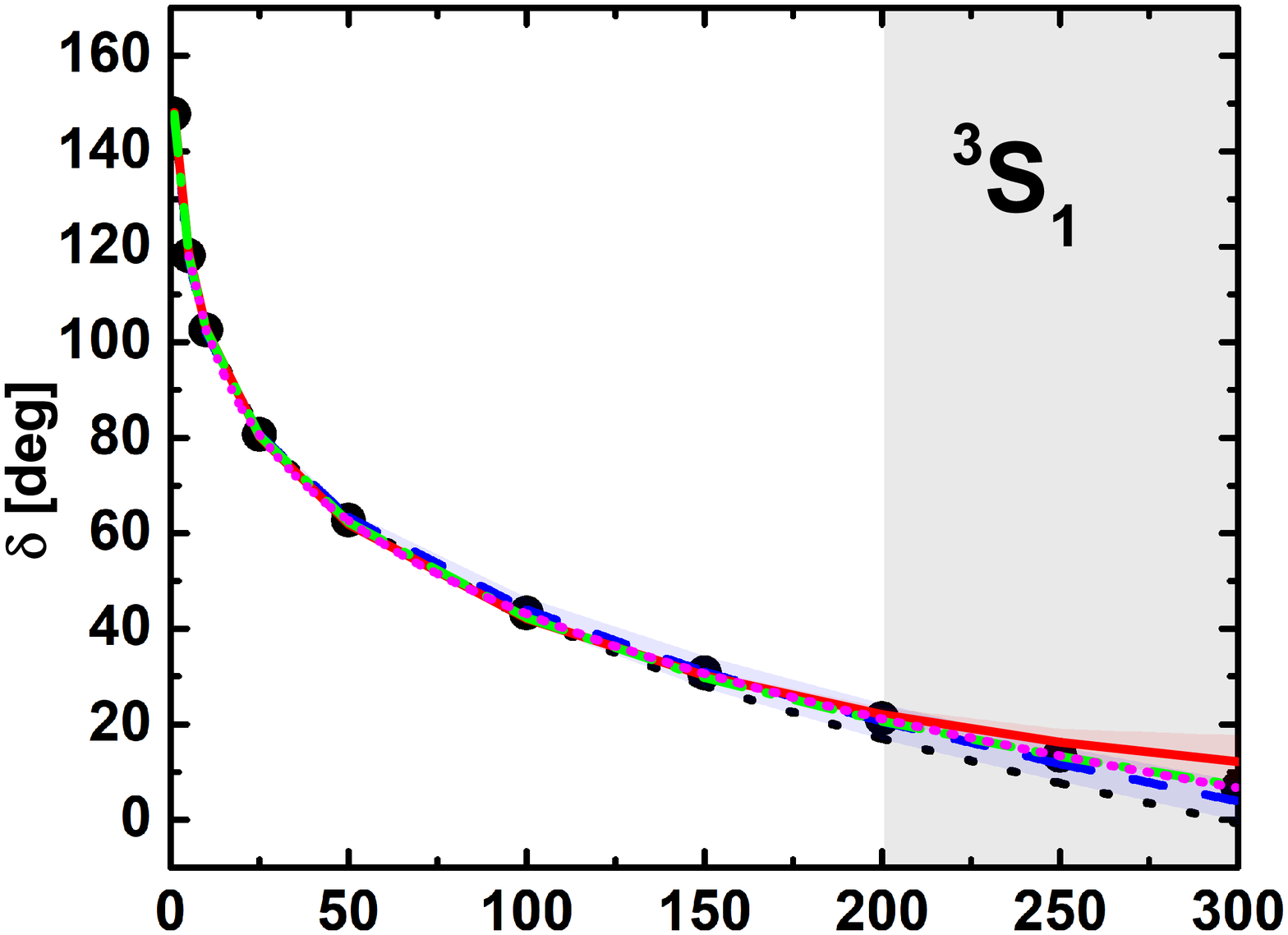}\hspace{-13mm}
\includegraphics[width=0.36\textwidth]{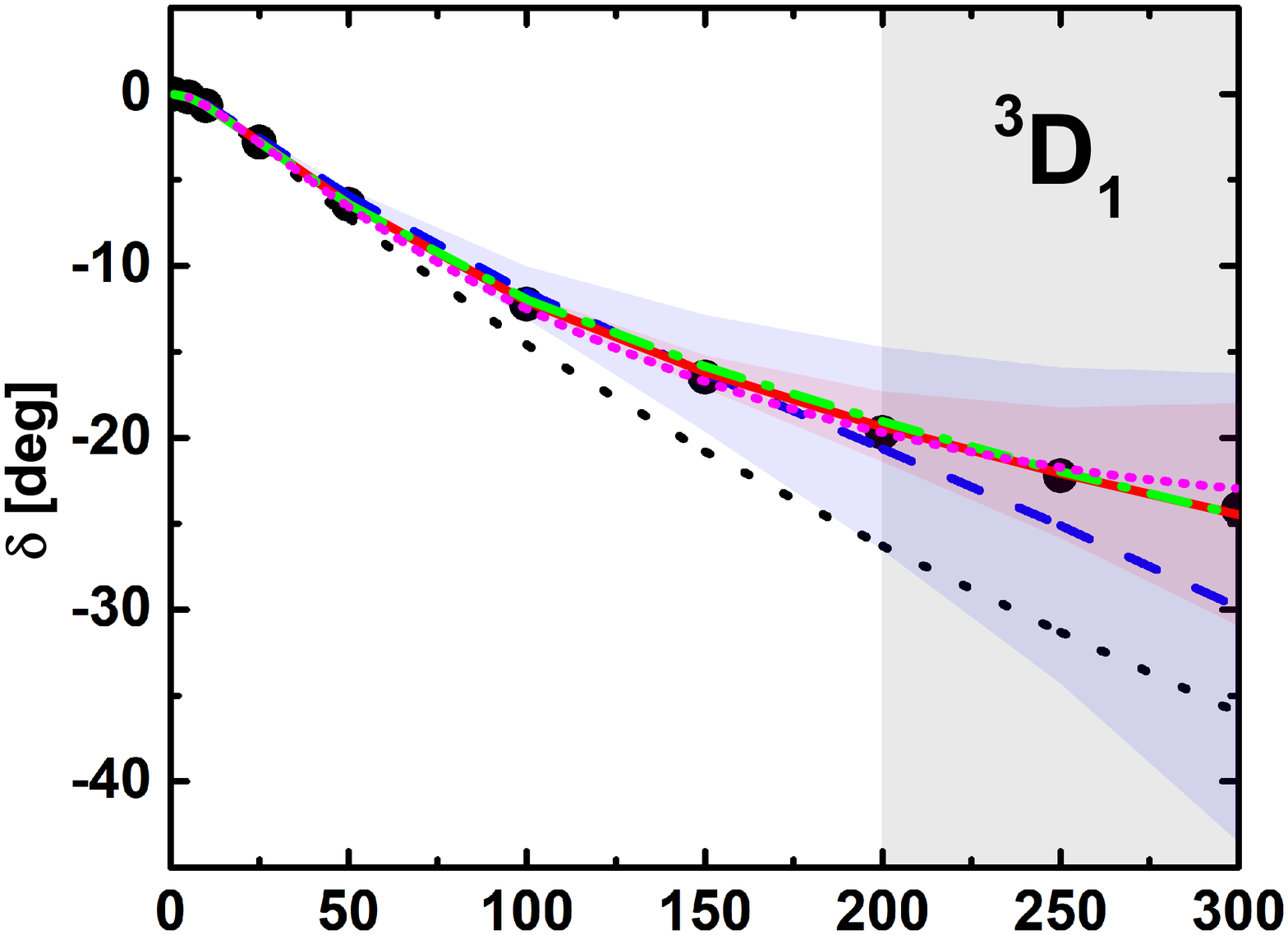}\hspace{-13mm}
\includegraphics[width=0.36\textwidth]{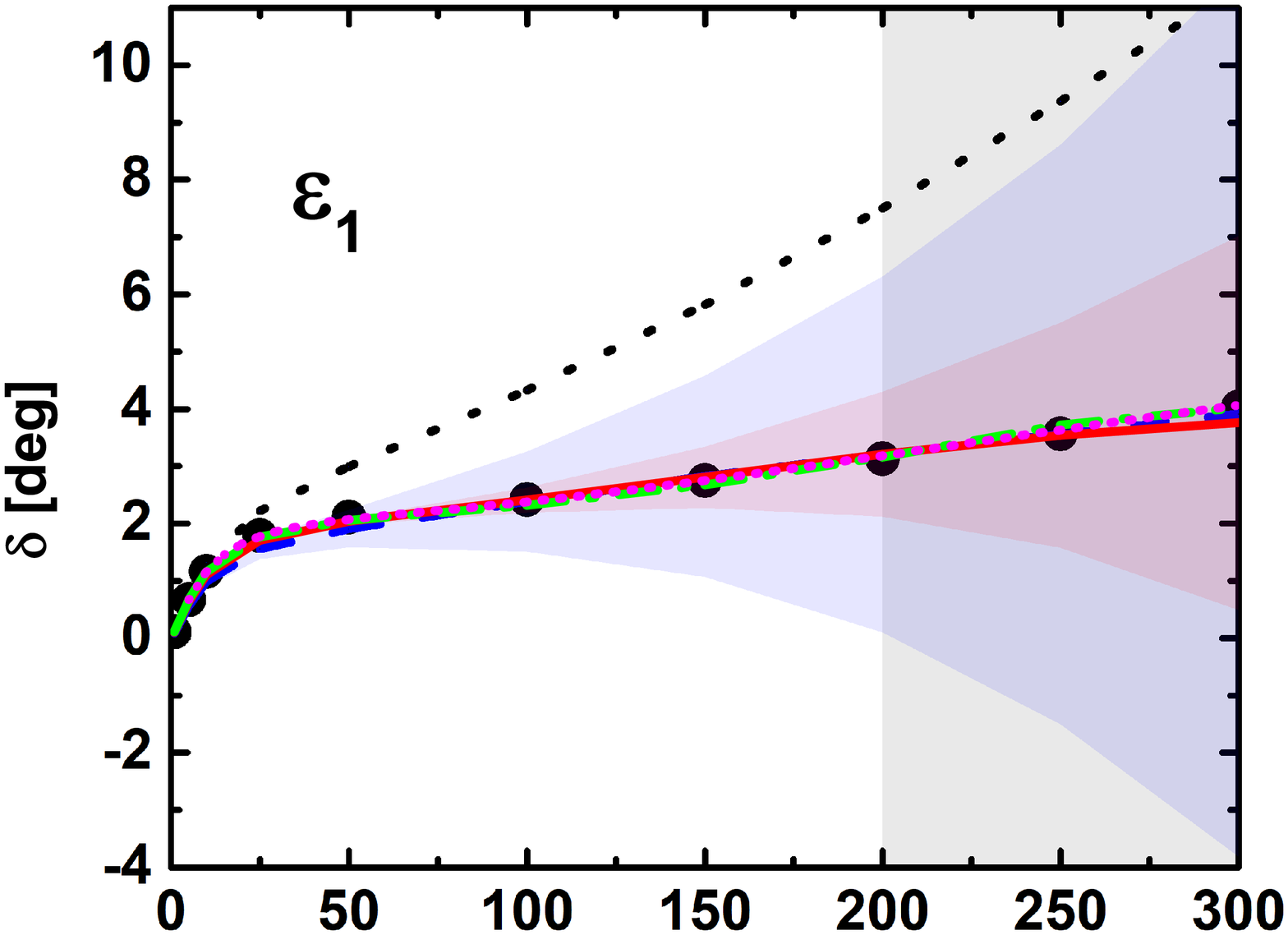}\\ \vspace{-9mm}
\includegraphics[width=0.36\textwidth]{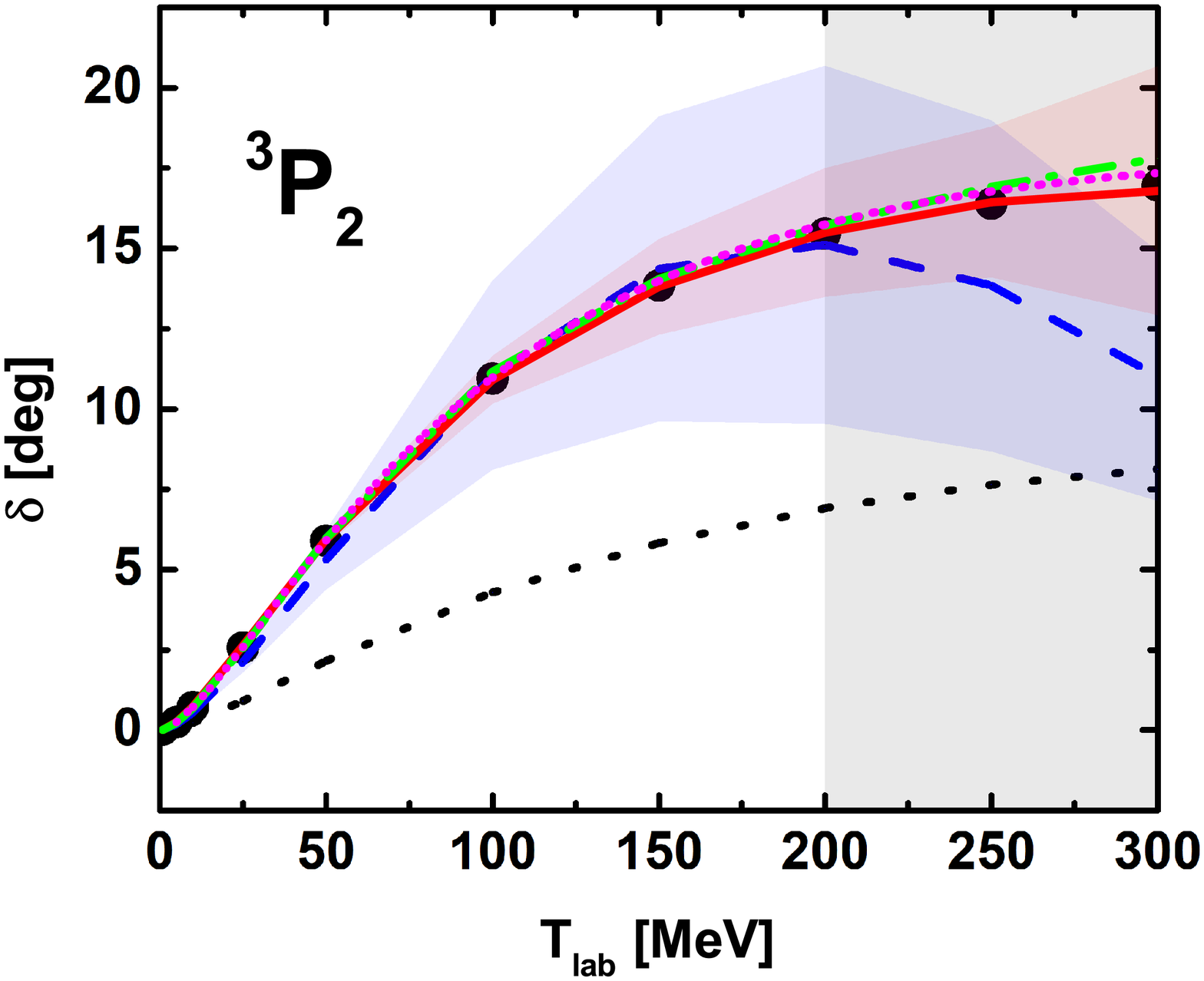}\hspace{-13mm}
\includegraphics[width=0.36\textwidth]{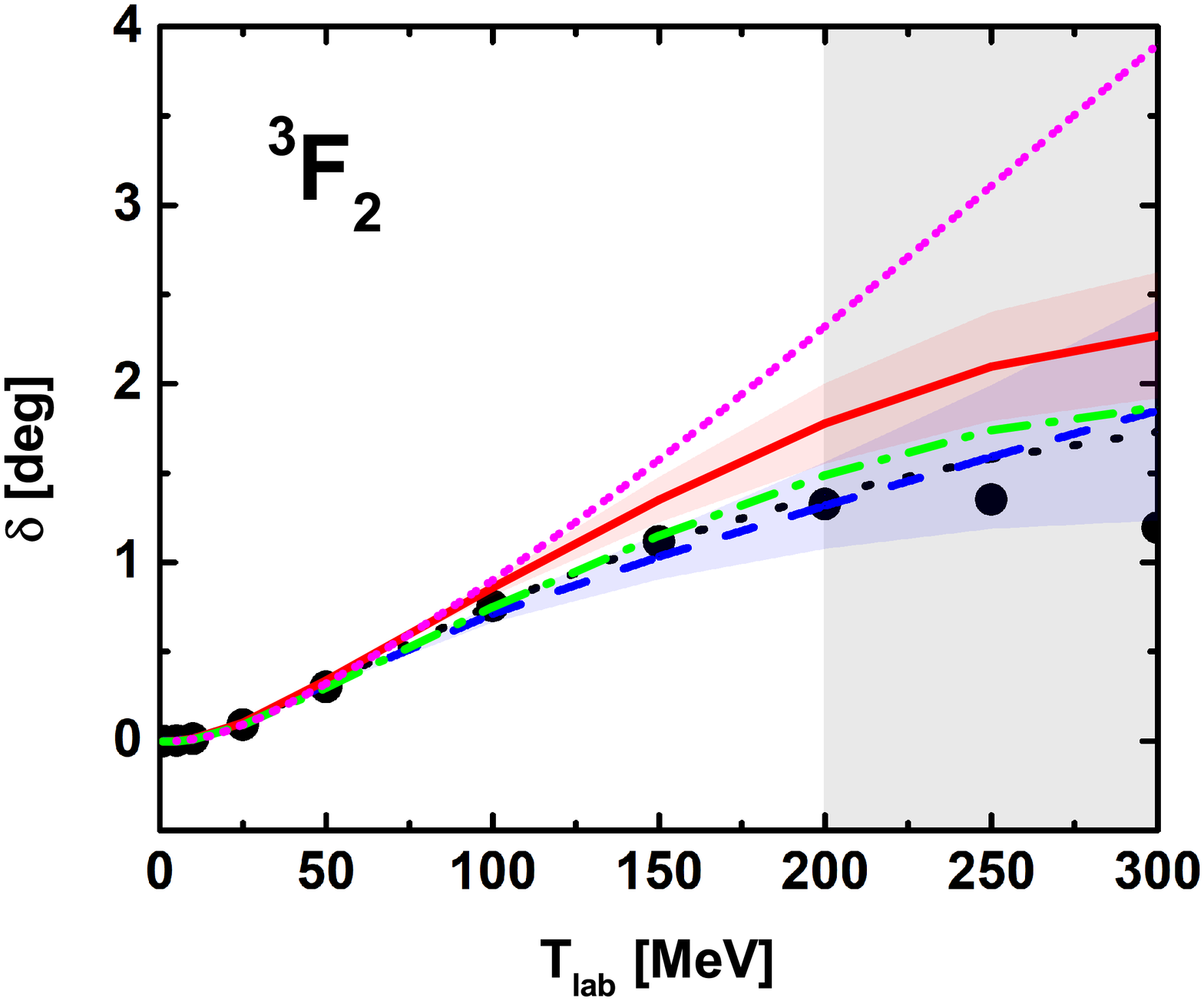}\hspace{-13mm}
\includegraphics[width=0.36\textwidth]{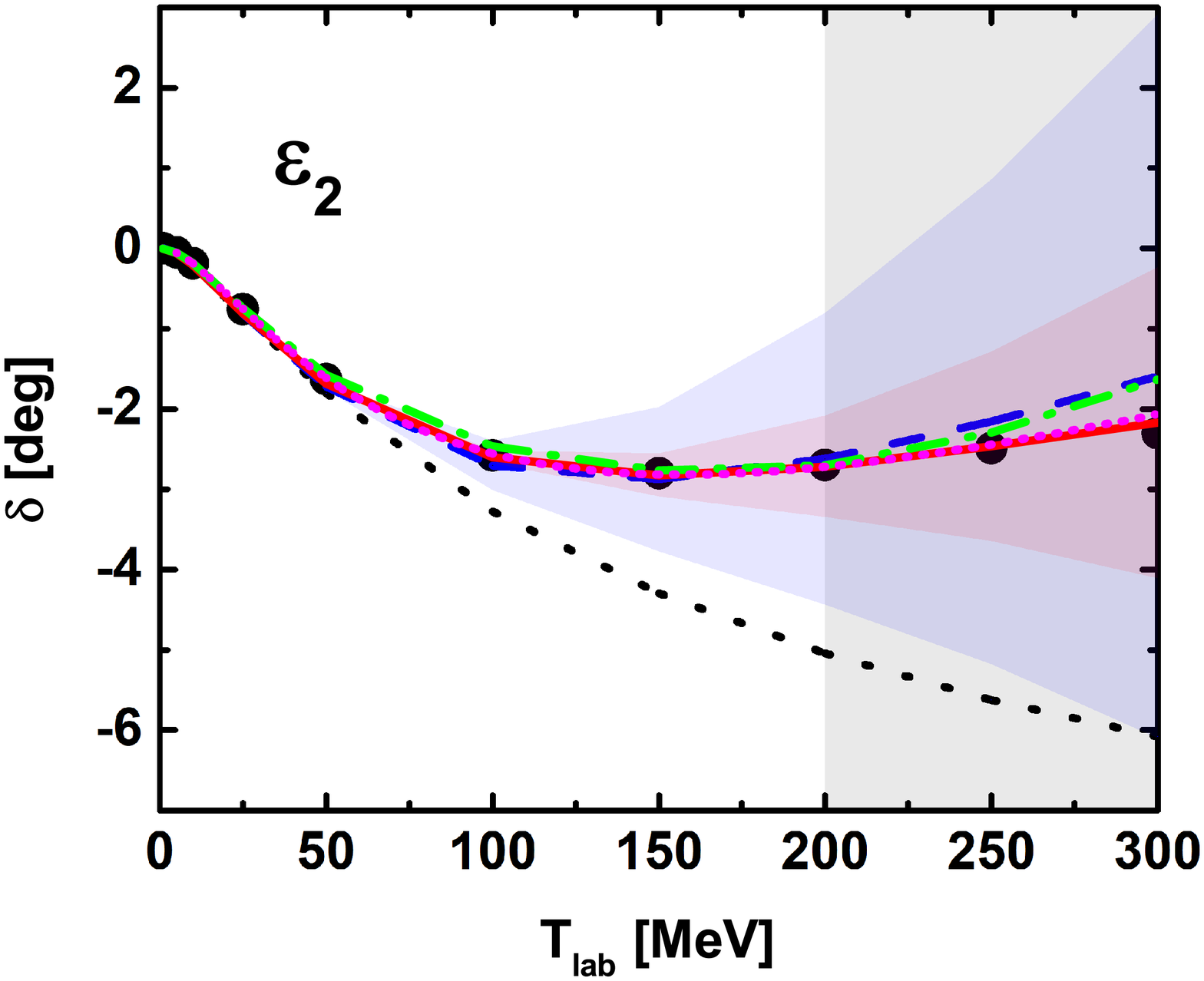}
\caption{$NN$ phaseshifts for partial waves with $J\leq 2$. The red solid lines denote the relativistic NNLO results obtained with a cutoff of $\Lambda=0.9$ GeV and the blue dashed lines denote the relativistic NLO  results obtained with a smaller cutoff of $\Lambda=0.6$ GeV. The corresponding bands represent the uncertainties for a DoB level of 68\%. For comparison, we also show the LO relativistic results (black dotted lines) obtained with a cutoff of $\Lambda=0.6$ GeV and the two sets of nonrelativistic N$^3$LO results  NR-N$^3$LO-Idaho ($\Lambda=0.5$ GeV, green dash-dotted lines)~\cite{Entem:2003ft,Machleidt:2011zz}  and NR-N$^3$LO-EKM (cutoff $=0.9$ fm, magenta short-dotted lines)~\cite{Epelbaum:2014efa,Epelbaum:2014sza}. The black dots denote  the PWA93 phaseshifts~\cite{Stoks:1993tb}. The shaded regions denote that those data are not fitted and the corresponding relativistic results are predictions.}
\label{fig:EX-uncertainties}
\end{figure*}

The so-obtained fitting results are shown in Fig.~\ref{fig:EX-uncertainties}, where the theoretical uncertainties are obtained via the Bayesian model for a DoB level of 68\%~\cite{Furnstahl:2015rha,Melendez:2017phj,Melendez:2019izc}.   For comparison, we also show the nonrelativistic N$^3$LO results obtained with different strategies for regularizing chiral potentials from Refs.~\cite{Entem:2003ft,Machleidt:2011zz} and Refs.~\cite{Epelbaum:2014efa,Epelbaum:2014sza} which are denoted as NR-N$^3$LO-Idaho and  NR-N$^3$LO-EKM, respectively.

 \begin{figure*}[htbp]
\centering
\includegraphics[width=0.35\textwidth]{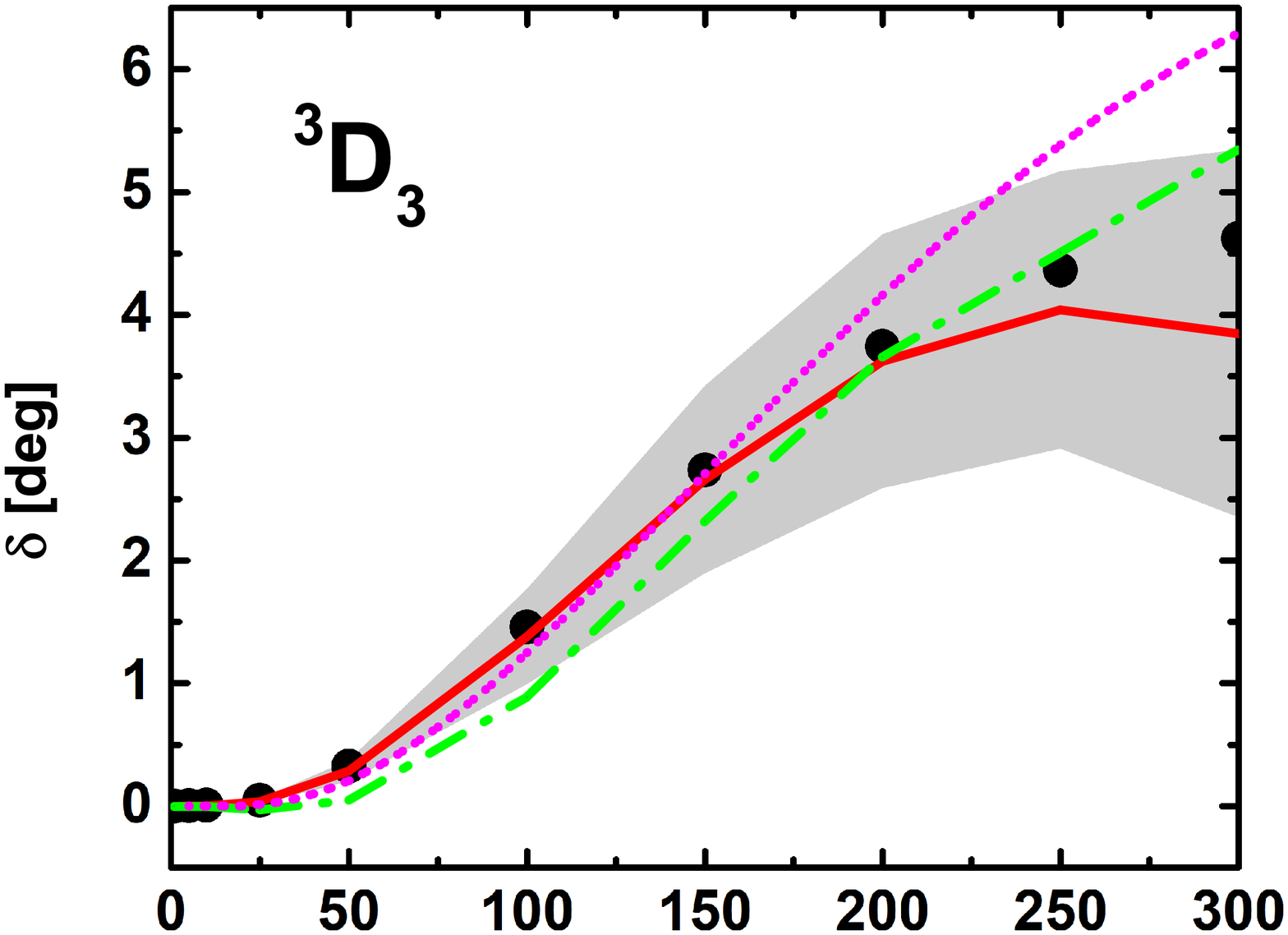}\hspace{-13mm}
\includegraphics[width=0.35\textwidth]{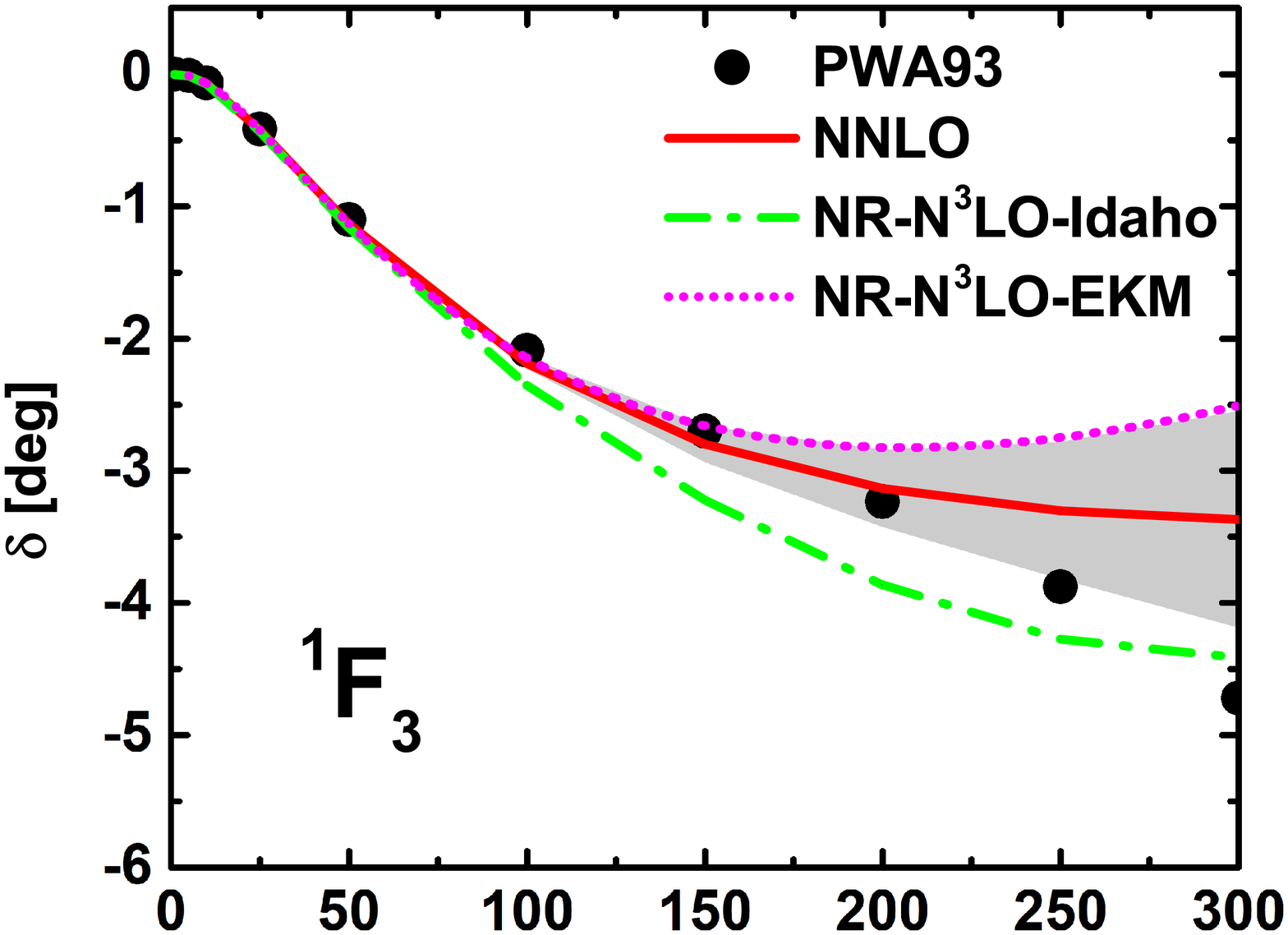}\hspace{-13mm}
\includegraphics[width=0.35\textwidth]{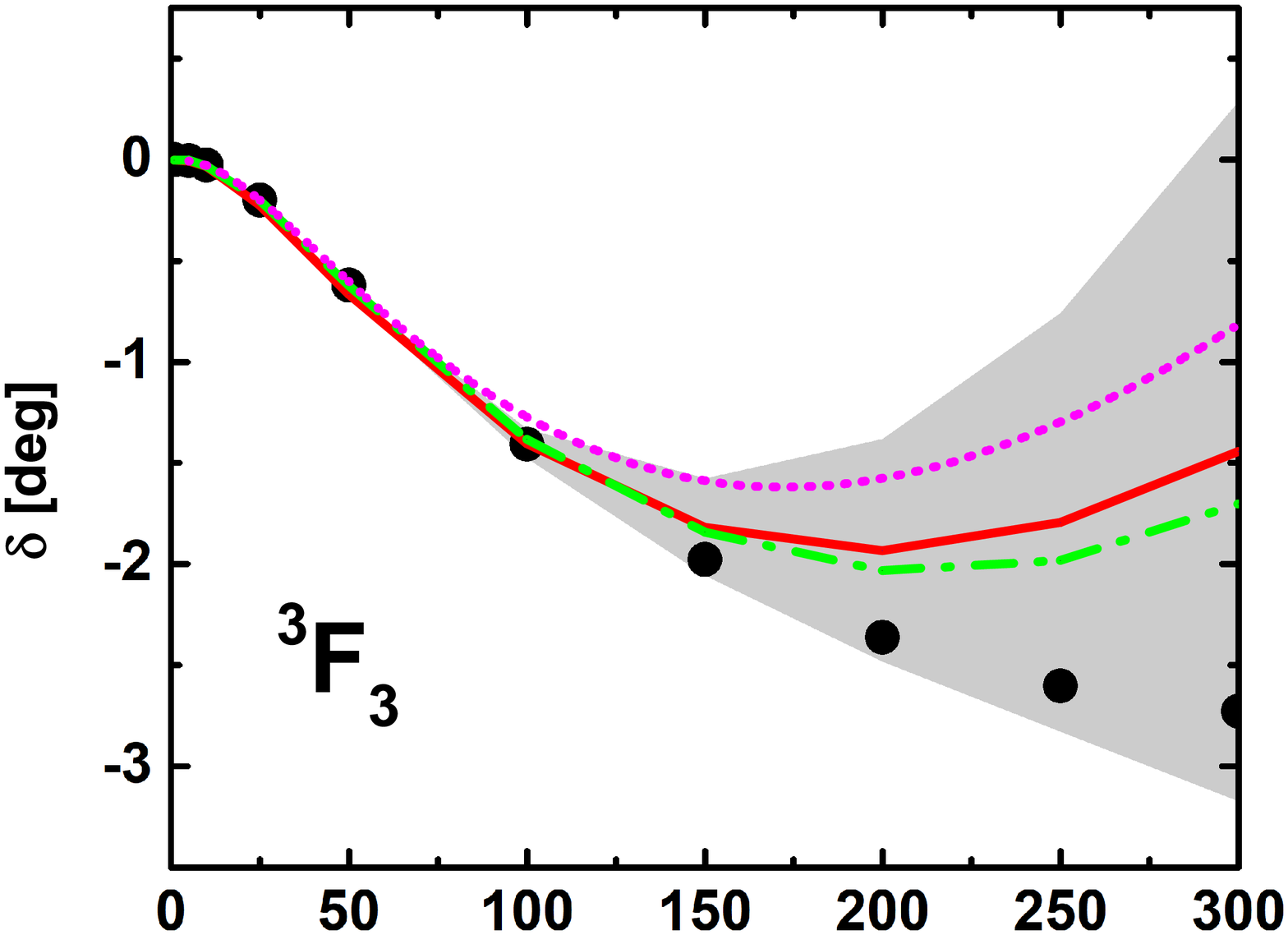}\\ \vspace{-6mm}
\includegraphics[width=0.35\textwidth]{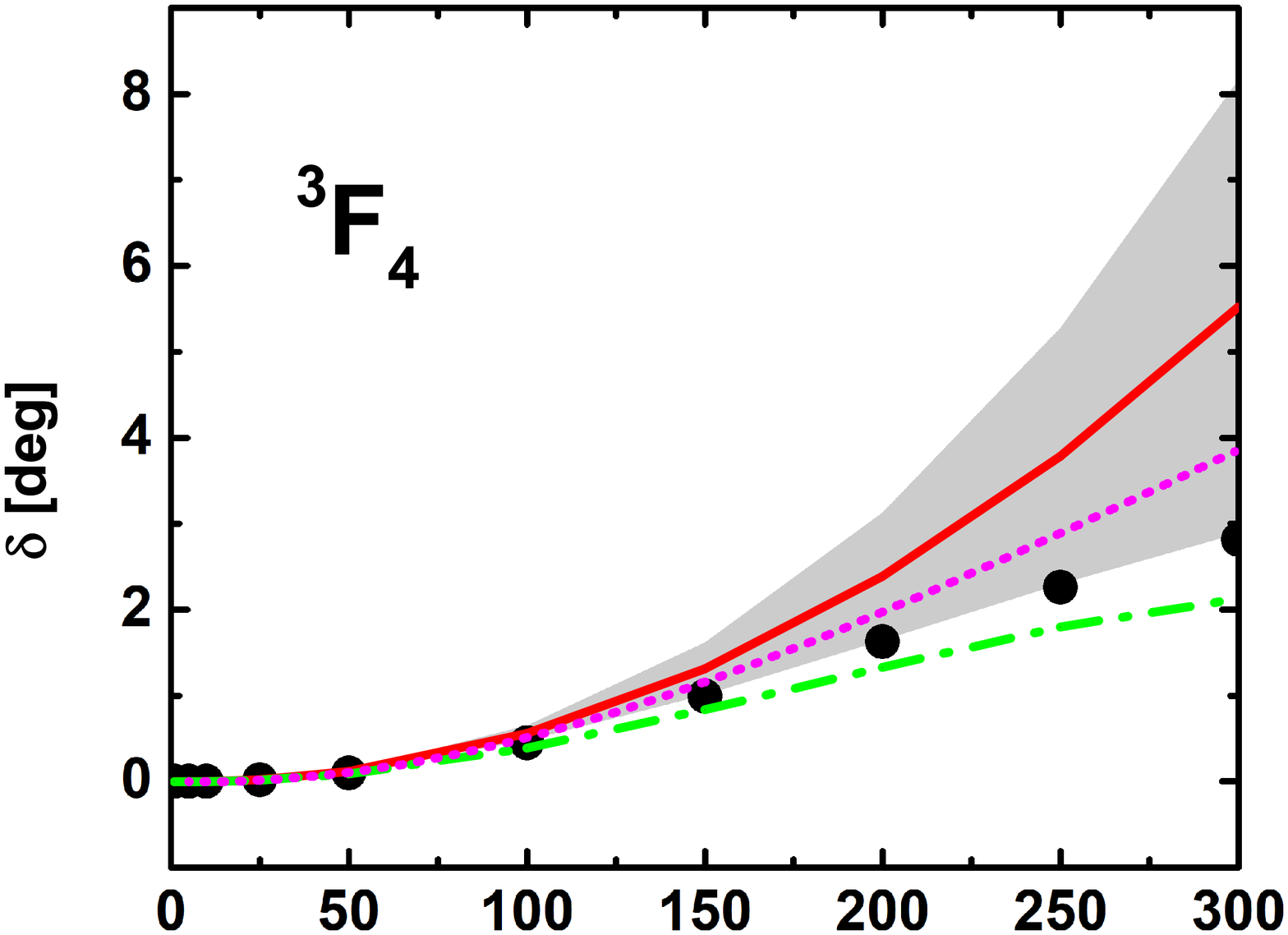}\hspace{-13mm}
\includegraphics[width=0.35\textwidth]{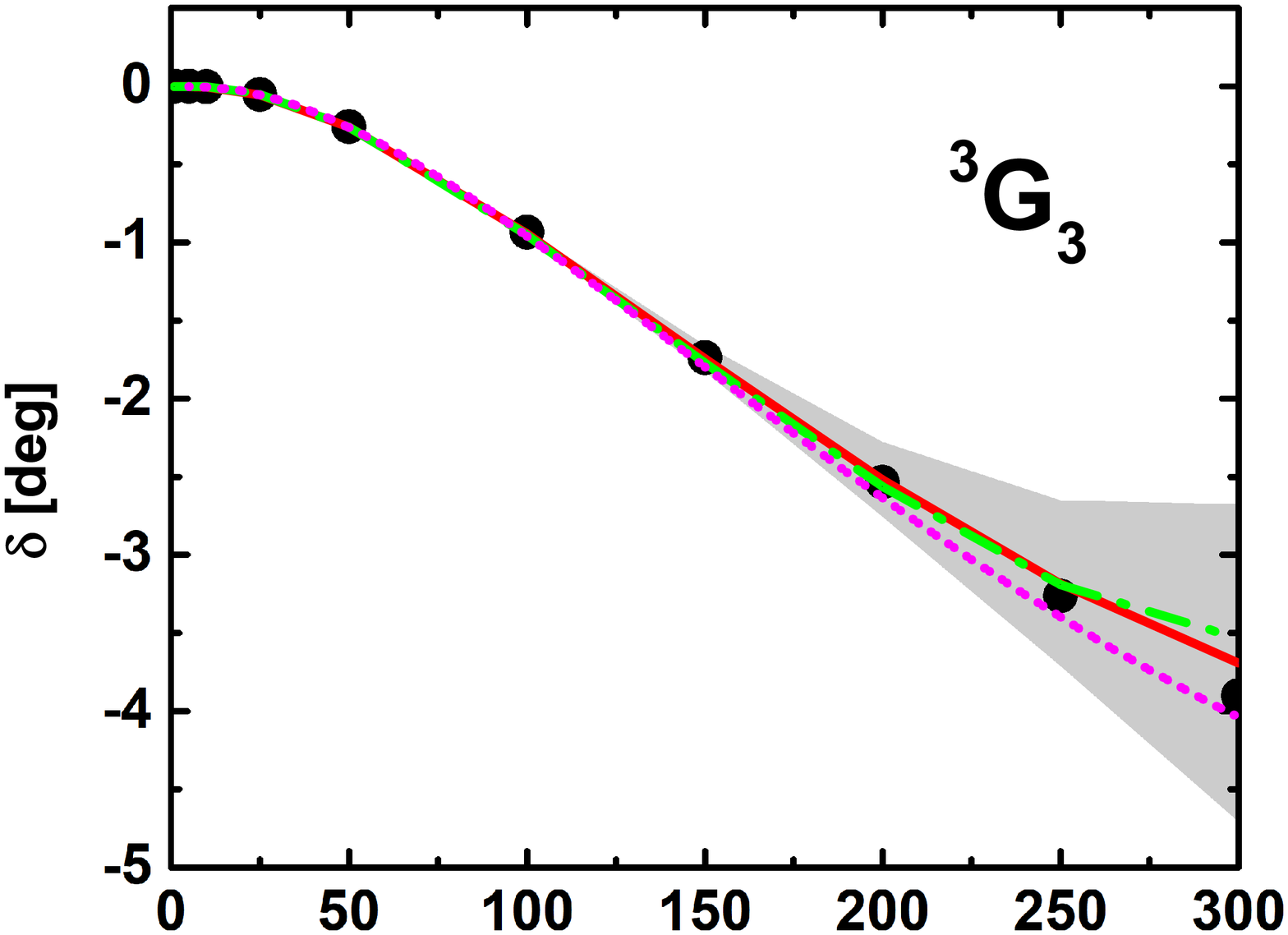}\hspace{-13mm}
\includegraphics[width=0.35\textwidth]{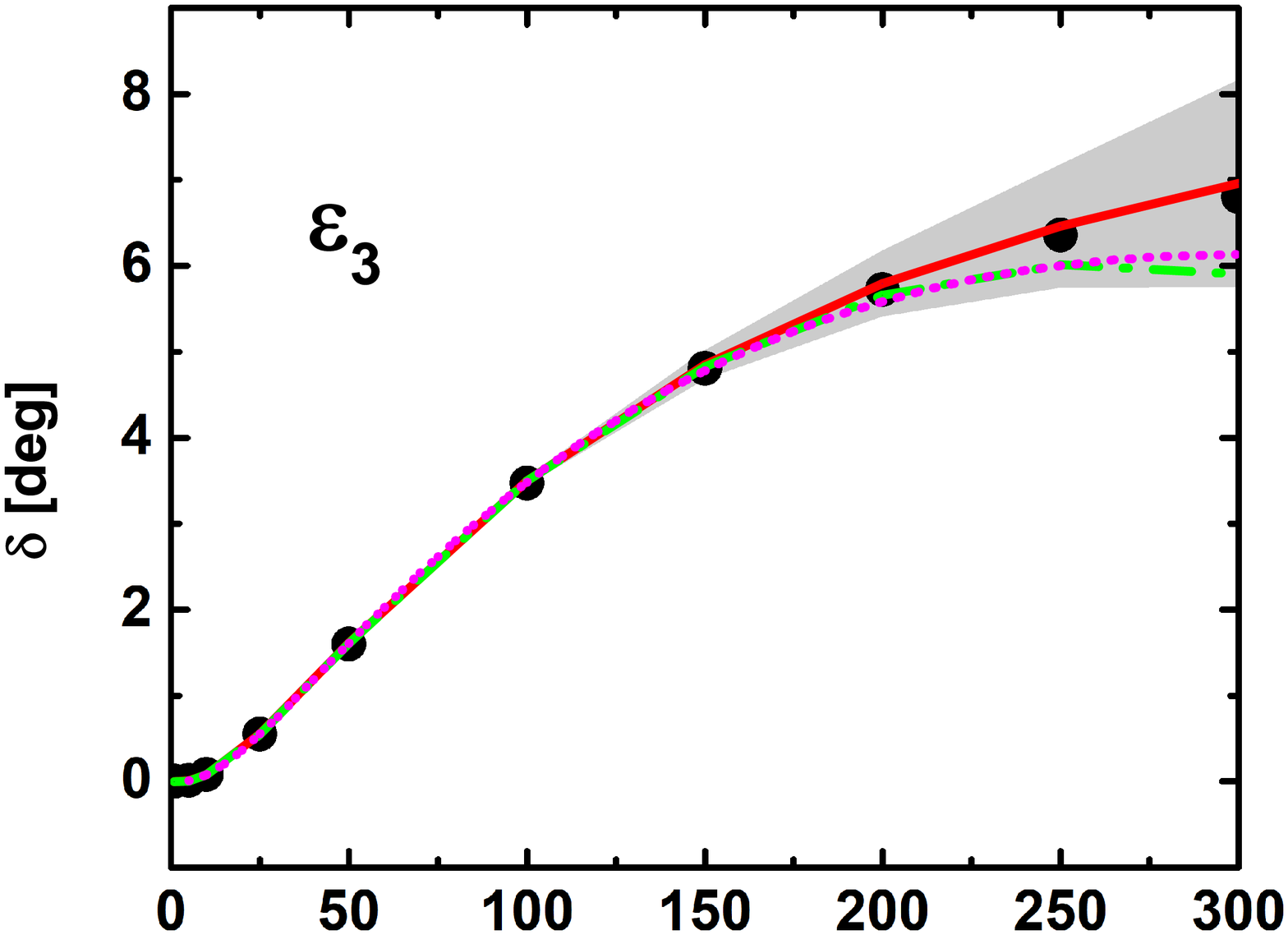}\\ \vspace{-6mm}
\includegraphics[width=0.35\textwidth]{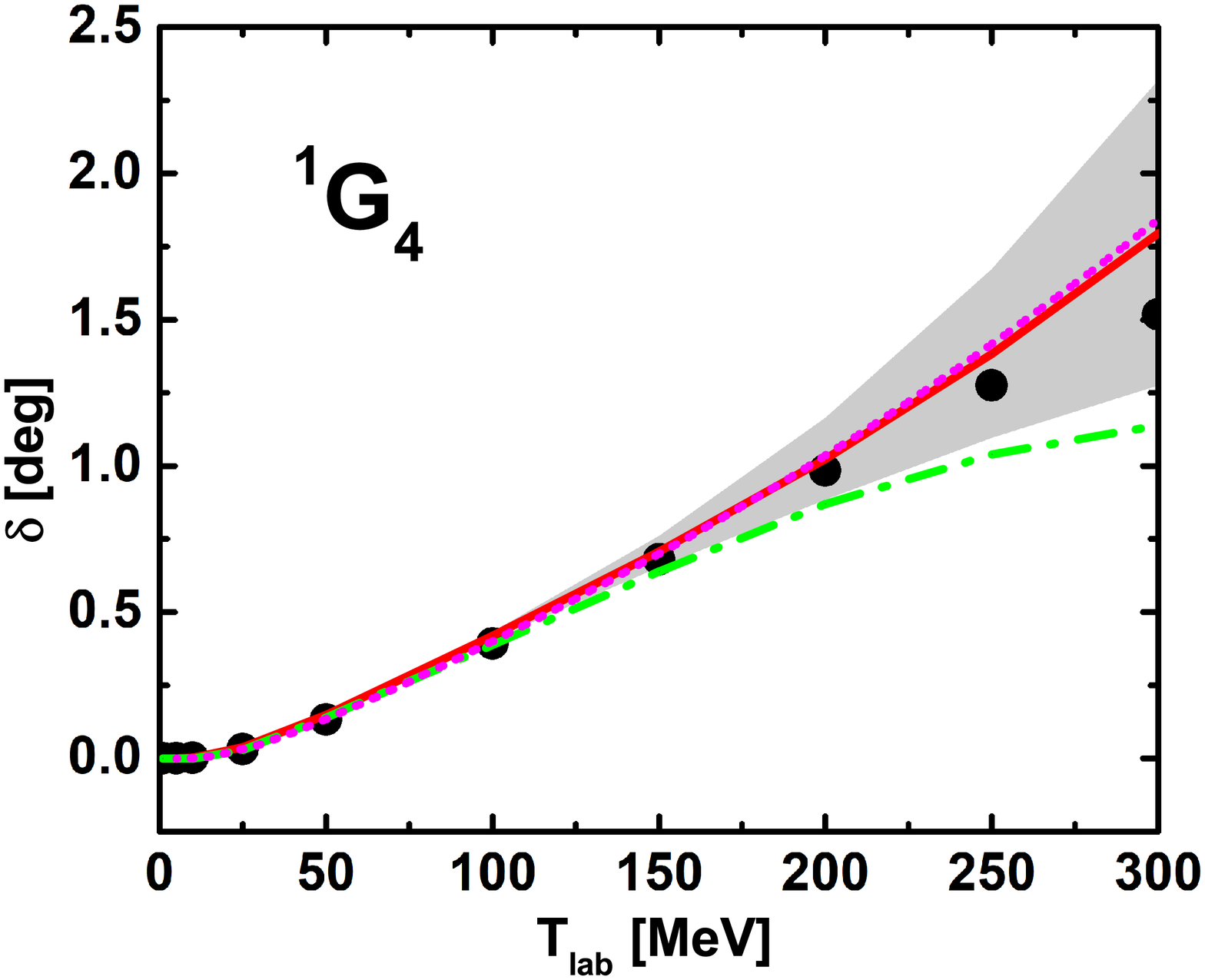}\hspace{-13mm}
\includegraphics[width=0.35\textwidth]{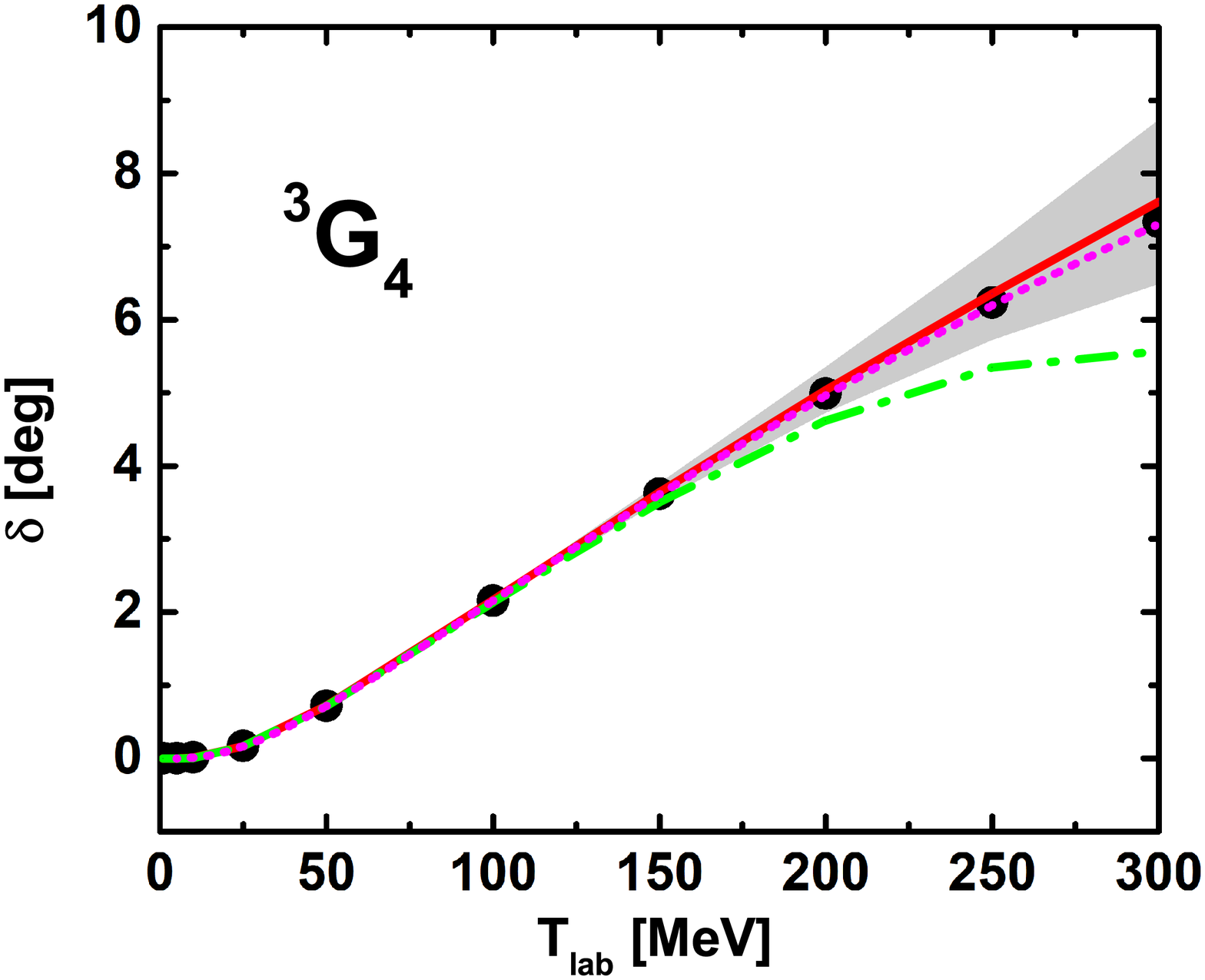}\hspace{-13mm}
\caption{Same as Fig.~\ref{fig:EX-uncertainties} but for peripheral partial waves with $J\le4$ and $L\le4$. Note that for these partial waves, the chiral results are predictions and independent from the partial wave analyses. Note that we do not show the LO and NLO relativistic chiral results for these peripheral partial waves.  }
\label{fig:High}
\end{figure*}

 First we notice that the NLO and NNLO relativistic results describe the $np$ phaseshifts very well up to $T_\mathrm{lab}=200$ MeV, at a level similar to the nonrelativistic N$^3$LO results. Particularly interesting is that the NLO and the NNLO results also agree well with each other for $T_\mathrm{lab}\le 200$ MeV, while the NNLO results are in better agreement the PWA93 data for larger kinetic energies. This demonstrates that the chiral series converge well. On the other hand, for $^3F_2$, the NLO results are better, which can be attributed to  the compromise that one has to make to fit all the $J=2$ partial waves with five LECs to balance the large contributions of subleading TPE. It can be improved once the correlation between the $D$-waves with $J=2$ and $^3P_2$-$^3F_2$ are removed, i.e., the $D$-waves and $^3P_2$-$^3F_2$ are fitted separately or the cutoff is slightly modified. We note that in obtaining the  NR-N$^3$LO-Idaho results, the phaseshifts of this channel were lowered by a careful fine-tuning of $c_2$ and $c_4$~\cite{Entem:2003ft}.

In Fig.~\ref{fig:High}, we compare the NR-N$^3$LO-Idaho~\cite{Entem:2003ft,Machleidt:2011zz} and the NR-N$^3$LO-EKM~\cite{Epelbaum:2014efa,Epelbaum:2014sza} chiral forces with the relativistic NNLO chiral force for peripheral partial waves with $J\le 4$ and $L\le4$. Higher partial waves are not explicitly shown  since for them the one-pion exchange contribution plays the dominant role.  Clearly for these partial waves, the relativistic NNLO  results are as good as or even slightly better than the nonrelativistic N$^3$LO results for $T_{\rm{lab}}\le200$ MeV except for $^3F_4$. For  $^3D_3$ and $^1F_3$, our relativistic NNLO results and those NR-N$^3$LO-EKM are not able to describe well the high-momentum data, while the NR-N$^3$LO-Idaho results  miss the data for $T_{\rm{lab}}\in$ [100,200] MeV. For $^3F_3$, no results can reproduce the behavior above $T_{\rm{lab}}=200$ MeV but the NR-N$^3$LO-Idaho results are slightly better. For $^3F_4$,  the subleading TPE is strong such that it shifts the relativistic results well above the PWA93 phaseshifts, while the NR-N$^3$LO-Idaho results behave much better. For the $G$-waves and the mixing angle $\epsilon_3$, all three results are in good agreement with the empirical data below $T_{\rm{lab}}=200$ MeV, while the NR-N$^3$LO-Idaho results tend to yield smaller values at higher energies.

\section{Summary and outlook}
To summarize, we constructed an accurate relativistic chiral nucleon-nucleon interaction up to the next-to-next-to-leading order in covariant baryon chiral perturbation theory and we obtained a good description of the PWA93 phaseshifts. The next-to-leading order (NLO) and the NNLO results agree well with each other for $T_\mathrm{lab}\le 200$ MeV, while at higher energies the NNLO results agree better with the PWA93 phaseshifts. This demonstrated the  convergence of the covariant chiral expansions. Given the quality already achieved in describing the $np$ phaseshifts, the NNLO relativistic chiral $NN$ interaction  provides the much wanted inputs for relativistic ab initio nuclear structure and reaction studies. In the future,  it will be interesting to extend the present study to the $u,d,s$ flavor sector and construct accurate hyperon-nucleon/hyperon interactions that are of great relevance for studies of neutron stars. Furthermore, one can study the antinucleon-nucleon interaction in the covariant framework as well. Such works are in progress.

\section{Acknowledgements}

This work is supported in part by the National Natural Science Foundation of China under Grants No.11735003, No.11975041,  and No.11961141004. Jun-Xu Lu acknowledges support from the National Natural Science Foundation of China under Grants No.12105006.

\end{document}